\documentstyle[12pt]{book}
\begin{document}

\sloppy
\raggedbottom

\chapter* 
{THE BERRY PARADOX}
\markright
{The Berry Paradox}
\addcontentsline{toc}{chapter}
{The Berry paradox}

\section*{G. J. Chaitin,
IBM Research Division,
P.~O. Box 704, Yorktown Heights, NY 10598, USA,
chaitin @ watson.ibm.com}
\section*{}

{\it
Lecture given Wednesday 27 October 1993 at a Physics -- Computer
Science Colloquium at the University of New Mexico.  The lecture was
videotaped; this is an edited transcript.  It also incorporates
remarks made at the Limits to Scientific Knowledge meeting held at the
Santa Fe Institute 24--26 May 1994.
}

In early 1974, I was visiting the Watson Research Center and I got the
idea of calling G\"odel on the phone.  I picked up the phone and
called and G\"odel answered the phone.  I said, ``Professor G\"odel,
I'm fascinated by your incompleteness theorem.  I have a new proof
based on the Berry paradox that I'd like to tell you about.''  G\"odel
said, ``It doesn't matter which paradox you use.''  He had used a
paradox called the liar paradox.  I said, ``Yes, but this suggests to
me an information-theoretic view of incompleteness that I would very
much like to tell you about and get your reaction.''  So G\"odel said,
``Send me one of your papers.  I'll take a look at it.  Call me again
in a few weeks and I'll see if I give you an appointment.''

I had had this idea in 1970, and it was 1974.  So far I had only
published brief abstracts.  Fortunately I had just gotten the galley
proofs of my first substantial paper on this subject.  I put these in
an envelope and mailed them to G\"odel.

I called G\"odel again and he gave me an appointment!  As you can
imagine I was delighted.  I figured out how to go to Princeton by
train.  The day arrived and it had snowed and there were a few inches
of snow everywhere.  This was certainly not going to stop me from
visiting G\"odel!  I was about to leave for the train and the phone
rang and it was G\"odel's secretary.  She said that G\"odel was very
careful about his health and because of the snow he wasn't coming to
the Institute that day and therefore my appointment was canceled.

And that's how I had two phone conversations with G\"odel but never
met him.  I never tried again.

I'd like to tell you what I would have told G\"odel.  What I wanted to
tell G\"odel is the difference between what you get when you study the
limits of mathematics the way G\"odel did using the paradox of the
liar, and what I get using the Berry paradox instead.

What is the paradox of the liar?  Well, the paradox of the liar is
\[
\begin{array}{l}
   \mbox{``This statement is false!''}
\end{array}
\]
Why is this a paradox?  What does ``false'' mean?  Well, ``false''
means ``does not correspond to reality.''  This statement says that it
is false.  If that doesn't correspond to reality, it must mean that
the statement is true, right?  On the other hand, if the statement is
true it means that what it says corresponds to reality.  But it says
that it is false.  Therefore the statement must be false.  So whether
you assume that it's true or false you then conclude the opposite!  So
this is the paradox of the liar.

Now let's look at the Berry paradox.  First of all, why ``Berry''?
Well it has nothing to do with fruit!  This paradox was published at
the beginning of this century by Bertrand Russell.  Now there's a
famous paradox which is called Russell's paradox and this is not it!
This is another paradox that he published.  I guess people felt that
if you just said the Russell paradox and there were two of them it
would be confusing.  And Bertrand Russell when he published this
paradox had a footnote saying that it was suggested to him by Mr G. G.
Berry.  So it ended up being called the Berry paradox even though it
was published by Russell.

Here is a version of the Berry paradox:
\[
\begin{array}{l}
   \mbox{``the first positive integer that cannot} \\
   \mbox{be specified in less than a billion words''.}
\end{array}
\]
This is a phrase in English that specifies a particular positive
integer.  Which positive integer?  Well, there are an infinity of
positive integers, but there are only a finite number of words in
English.  Therefore, if you have a billion words, there's only
going to be a finite number of possibilities.  But there's an infinite
number of positive integers.  Therefore most positive integers require
more than a billion words, and let's just take the first one.  But
wait a second.  By definition this integer is supposed to take a
billion words to specify, but I just specified it using much less than
a billion words!  That's the Berry paradox.

What does one do with these paradoxes?  Let's take a look again at the
liar paradox:
\[
\begin{array}{l}
   \mbox{``This statement is false!''}
\end{array}
\]
The first thing that G\"odel does is to change it from ``This
statement is false'' to ``This statement is unprovable'':
\[
\begin{array}{l}
   \mbox{``This statement is unprovable!''}
\end{array}
\]
What do we mean by ``unprovable''?

In order to be able to show that mathematical reasoning has limits
you've got to say very precisely what the axioms and methods of
reasoning are that you have in mind.  In other words, you have to
specify how mathematics is done with mathematical precision so that it
becomes a clear-cut question.  Hilbert put it this way: The rules
should be so clear, that if somebody gives you what they claim is a
proof, there is a mechanical procedure that will check whether the
proof is correct or not, whether it obeys the rules or not.  This
proof-checking algorithm is the heart of this notion of a completely
formal axiomatic system.

So ``This statement is unprovable'' doesn't mean unprovable in a vague
way.  It means unprovable when you have in mind a specific formal
axiomatic system {\sl FAS\/} with its mechanical proof-checking
algorithm.  So there is a subscript:
\[
\begin{array}{l}
   \mbox{``This statement is unprovable${}_{\it FAS}$!''}
\end{array}
\]

And the particular formal axiomatic system that G\"odel was interested
in dealt with 1, 2, 3, 4, 5, and addition and multiplication, that was
what it was about.  Now what happens with ``This statement is
unprovable''?  Remember the liar paradox:
\[
\begin{array}{l}
   \mbox{``This statement is false!''} \\
\end{array}
\]
But here
\[
\begin{array}{l}
   \mbox{``This statement is unprovable${}_{\it FAS}$!''}
\end{array}
\]
the paradox disappears and we get a theorem.  We get incompleteness,
in fact.  Why?

Consider ``This statement is unprovable''.  There are two
possibilities: either it's provable or it's unprovable.

If ``This statement is unprovable'' turns out to be unprovable within
the formal axiomatic system, that means that the formal axiomatic
system is incomplete.  Because if ``This statement is unprovable'' is
unprovable, then it's a true statement.  Then there's something true
that's unprovable which means that the system is incomplete.  So that
would be bad.

What about the other possibility?  What if
\[
\begin{array}{l}
   \mbox{``This statement is unprovable${}_{\it FAS}$!''}
\end{array}
\]
is provable?  That's even worse.  Because if this statement is
provable
\[
\begin{array}{l}
   \mbox{``This statement is unprovable${}_{\it FAS}$!''}
\end{array}
\]
and it says of itself that it's unprovable, then we're proving
something that's false.

So G\"odel's incompleteness result is that if you assume that only
true theorems are provable, then this
\[
\begin{array}{l}
   \mbox{``This statement is unprovable${}_{\it FAS}$!''}
\end{array}
\]
is an example of a statement that is true but unprovable.

But wait a second, how can a statement deny that it is provable?  In
what branch of mathematics does one encounter such statements?
G\"odel cleverly converts this
\[
\begin{array}{l}
   \mbox{``This statement is unprovable${}_{\it FAS}$!''}
\end{array}
\]
into an arithmetical statement, a statement that only involves 1, 2,
3, 4, 5, and addition and multiplication.  How does he do this?

The idea is called g\"odel numbering.  We all know that a string of
characters can also be thought of as a number.  Characters are either
8 or 16 bits in binary.  Therefore, a string of $N$ characters is
either $8N$ or $16N$ bits, and it is also the base-two notation for a
large positive integer.  Thus every mathematical statement in this
formal axiomatic system
\[
\begin{array}{l}
   \mbox{``This statement is unprovable${}_{FAS\longleftarrow}$!''}
\end{array}
\]
is also a number.  And a proof, which is a sequence of steps, is also
a long character string, and therefore is also a number.  Then you can
define this very funny numerical relationship between two numbers $X$
and $Y$ which is that $X$ is the g\"odel number of a proof of the
statement whose g\"odel number is $Y$.  Thus
\[
\begin{array}{l}
   \mbox{``This statement is unprovable${}_{\it FAS}$!''}
\end{array}
\]
ends up looking like a very complicated numerical statement.

There is another serious difficulty.  How can this statement refer to
itself?  Well you can't directly put the g\"odel number of this
statement inside this statement, it's too big to fit!  But you can do
it indirectly.  This is how G\"odel does it: The statement doesn't
refer to itself directly.  It says that if you perform a certain
procedure to calculate a number, this is the g\"odel number of a
statement which cannot be proved.  And it turns out that the number
you calculate is precisely the g\"odel number of the entire statement
\[
\begin{array}{l}
   \mbox{``This statement is unprovable${}_{\it FAS}$!''}
\end{array}
\]
That is how G\"odel proves his incompleteness theorem.

What happens if you start with this
\[
\begin{array}{l}
   \mbox{``the first positive integer that cannot} \\
   \mbox{be specified in less than a billion words''}
\end{array}
\]
instead?  Everything has a rather different flavor.  Let's see why.

The first problem we've got here is what does it mean to specify a
number using words in English?---this is very vague.  So instead let's
use a computer.  Pick a standard general-purpose computer, in other
words, pick a universal Turing machine ({\sl UTM\/}).  Now the way you specify
a number is with a computer program.  When you run this computer
program on your {\sl UTM} it prints out this number and halts.  So a
program is said to specify a number, a positive integer, if you start
the program running on your standard {\sl UTM,} and after a finite amount of
time it prints out one and only one great big positive integer and it
says ``I'm finished'' and halts.

Now it's not English text measured in words, it's computer programs
measured in bits.  This is what we get.  It's
\[
\begin{array}{l}
   \mbox{``the first positive integer that cannot} \\
   \mbox{be specified${}_{\it UTM}$ by a computer program} \\
   \mbox{with less than a billion bits''.}
\end{array}
\]
By the way the computer program must be self-contained.  If it has
any data, the data is included in the program as a constant.

Next we have to do what G\"odel did when he changed ``This statement
is false'' into ``This statement is unprovable.''  So now it's
\[
\begin{array}{l}
   \mbox{``the first positive integer that can be proved${}_{\it FAS}$} \\
   \mbox{to have the property that it cannot} \\
   \mbox{be specified${}_{\it UTM}$ by a computer program} \\
   \mbox{with less than a billion bits''.}
\end{array}
\]
And to make things clearer let's replace ``a billion bits'' by ``$N$ bits''.
So we get:
\[
\begin{array}{l}
   \mbox{``the first positive integer that can be proved${}_{\it FAS}$} \\
   \mbox{to have the property that it cannot} \\
   \mbox{be specified${}_{\it UTM}$ by a computer program} \\
   \mbox{with less than $N$ bits''.}
\end{array}
\]

The interesting fact is that there is a computer program
\[
    \log_2 N + c_{\it FAS}
\]
bits long for calculating this number that supposedly cannot be calculated
by any program that is less than $N$ bits long. And
\[
    \log_2 N + c_{\it FAS}
\]
is much much smaller than $N$ for all sufficiently large $N$!  Thus
for such $N$ our {\sl FAS\/} cannot enable us to exhibit any numbers
that require programs more than $N$ bits long.  This is my
information-theoretic incompleteness result that I wanted to discuss
with G\"odel.

Why is there a program that is
\[
    \log_2 N + c_{\it FAS}
\]
bits long for calculating
\[
\begin{array}{l}
   \mbox{``the first positive integer that can be proved${}_{\it FAS}$} \\
   \mbox{to have the property that it cannot} \\
   \mbox{be specified${}_{\it UTM}$ by a computer program} \\
   \mbox{with less than $N$ bits'' ?}
\end{array}
\]
Well here is how you do it.

You start running through all possible proofs in the formal axiomatic
system in size order.  You apply the proof-checking algorithm to each
proof.  And after filtering out all the invalid proofs, you search for
the first proof that a particular positive integer requires at least
an $N$-bit program.

The algorithm that I've just described is very slow but it is very
simple.  It's basically just the proof-checking algorithm, which is
$c_{\it FAS}$ bits long, and the number $N$, which is $\log_2 N$ bits
long.  So the total number of bits is, as was claimed, just
\[
    \log_2 N + c_{\it FAS} .
\]
That concludes the proof of my incompleteness result that I wanted to
discuss with G\"odel.

Over the years I've continued to do research on my
information-theoretic approach to incompleteness.  Here are the three
most dramatic results that I've obtained:

\begin{itemize}

\item[1)] Call a program ``elegant'' if no smaller program produces
the same output.  You can't prove that a program is elegant.  More
precisely, $N$ bits of axioms are needed to prove that an $N$-bit
program is elegant.

\item[2)] Consider the binary representation of the halting
probability $\Omega$.  $\Omega$ is the probability that a program
chosen at random halts.  You can't prove what the bits of $\Omega$
are.  More precisely, $N$ bits of axioms are needed to determine $N$
bits of $\Omega$.

\item[3)] I have constructed a perverse algebraic equation
\[
   P(K,X,Y,Z,\ldots) = 0.
\]
Vary the parameter $K$ and ask whether this equation has finitely or
infinitely many whole-number solutions.  In each case this turns out
to be equivalent to determining individual bits of $\Omega$.
Therefore $N$ bits of axioms are needed to be able to settle $N$
cases.

\end{itemize}

These striking examples show that sometimes you have to put more into
a set of axioms in order to get more out.  (2) and (3) are extreme
cases.  They are accidental mathematical assertions that are true for
no reason.  In other words, the questions considered in (2) and (3)
are irreducible; essentially the only way to prove them is to add them
as new axioms.  Thus in this extreme case you get out of a set of
axioms only what you put in.

How do I prove these incompleteness results (1), (2) and (3)?  As
before, the basic idea is the paradox of ``the first positive integer
that cannot be specified in less than a billion words.''  For (1) the
connection with the Berry paradox is obvious.  For (2) and (3) it was
obvious to me only in the case where one is talking about determining
the {\bf first} $N$ bits of $\Omega$.  In the case where the $N$ bits
of $\Omega$ are scattered about, my original proof of (2) and (3) (the
one given in my Cambridge University Press monograph) is decidedly not
along the lines of the Berry paradox.  But a few years later I was
happy to discover a new and more straight-forward proof of (2) and (3)
that is along the lines of the Berry paradox!

In addition to working on incompleteness, I have also devoted a great
deal of thought to the central idea that can be extracted from my
version of the Berry paradox, which is to define the program-size
complexity of something to be the size in bits of the smallest program
that calculates it.  I have developed a general theory dealing with
program-size complexity that I call algorithmic information theory
({\sl AIT\/}).

{\sl AIT\/} is an elegant theory of complexity, perhaps the most
developed of all such theories, but as von Neumann said, pure
mathematics is easy compared to the real world!  {\sl AIT\/} provides
the correct complexity concept for metamathematics, but it is not the
correct complexity concept for physics, biology, or economics.

Program-size complexity in {\sl AIT\/} is analogous to entropy in
statistical mechanics.  Just as thermodynamics gives limits on heat
engines, {\sl AIT\/} gives limits on formal axiomatic systems.

I have recently reformulated {\sl AIT.}

Up to now, the best version of {\sl AIT\/} studied the size of
programs in a computer programming language that was not actually
usable.  Now I obtain the correct program-size complexity measure from
a powerful and easy to use programming language.  This language is a
version of {\sl LISP,} and I have written an interpreter for it in
{\sl C.} A summary of this new work is available as IBM Research
Report RC 19553 ``The limits of mathematics,'' which I am expanding
into a book.

So this is what I would have liked to discuss with G\"odel, if I could
speak with him now.  Of course this is impossible!  But thank you very
much for giving me the opportunity to tell you about these ideas!

\section*{Questions for Future Research}

\begin{itemize}

\item Find questions in algebra, topology and geometry that are
equivalent to determining bits of $\Omega$.

\item What is an interesting or natural mathematical question?

\item How often is such a question independent of the usual axioms?
(I suspect the answer is ``Quite often!'')

\item Show that a classical open question in number theory such as
the Riemann hypothesis is independent of the usual axioms.  (I suspect
that this is often the case, but that it cannot be proven.)

\item Should we take incompleteness seriously or is it a red
herring?  (I believe that we should take incompleteness very seriously
indeed.)

\item Is mathematics quasi-empirical?  In other words, should
mathematics be done more like physics is done?  (I believe the answer
to both questions is ``Yes.'')

\end{itemize}

\section*{Bibliography}

{\bf Books:}

\begin{itemize}

\item
G. J. Chaitin,
{\it Information, Randomness \& Incompleteness,}
second edition, World Scientific, 1990.
Errata:
on page 26, line 25, ``quickly that''
should read ``quickly than'';
on page 31, line 19, ``Here one''
should read ``Here once'';
on page 55, line 17, ``RI, p.\ 35''
should read ``RI, 1962, p.\ 35'';
on page 85, line 14, ``1.\ The problem''
should read ``1.\ The Problem'';
on page 88, line 13, ``4.\ What is life?''
should read ``4.\ What is Life?'';
on page 108, line 13, ``the table in''
should read ``in the table in'';
on page 117, Theorem 2.3(q),
``$H_{C}(s,t)$''
should read ``$H_{C}(s/t)$'';
on page 134, line 7,
``$\# \{ n | H(n) \leq n \} \leq 2^{n}$''
should read
``$\# \{ k | H(k) \leq n \} \leq 2^{n}$'';
on page 274, bottom line, ``$n_{4p+4}$''
should read ``$n_{4p'+4}$''.

\item
G. J. Chaitin,
{\it Algorithmic Information Theory,}
fourth printing, Cambridge University Press, 1992.
Erratum:
on page 111, Theorem I0(q),
``$H_{C}(s,t)$''
should read ``$H_{C}(s/t)$''.

\item
G. J. Chaitin,
{\it Information-Theoretic Incompleteness,}
World Scientific, 1992.
Errata:
on page 67, line 25, ``are there are''
should read ``are there'';
on page 71, line 17, ``that case that''
should read ``the case that'';
on page 75, line 25, ``the the''
should read ``the'';
on page 75, line 31, ``$-\log_2p-\log_2q$''
should read ``$-p\log_2p-q\log_2q$'';
on page 95, line 22, ``This value of''
should read ``The value of'';
on page 98, line 34, ``they way they''
should read ``the way they'';
on page 99, line 16, ``exactly same''
should read ``exactly the same'';
on page 124, line 10, ``are there are''
should read ``are there''.

\end{itemize}
{\bf Recent Papers:}
\begin{itemize}

\item
G. J. Chaitin,
``On the number of $n$-bit strings with maximum complexity,''
{\it Applied Mathematics and Computation\/} {\bf 59} (1993), pp.\ 97--100.

\item
G. J. Chaitin,
``Randomness in arithmetic and the decline and fall of reductionism in
pure mathematics,''
{\it Bulletin of the European Association for Theoretical Computer Science,}
No.\ 50 (June 1993), pp.\ 314--328.

\item
G. J. Chaitin,
``Exhibiting randomness in arithmetic using Mathematica and C,''
{\it IBM Research Report RC-18946,} 94 pp., June 1993.

\item
G. J. Chaitin,
``The limits of mathematics---Course outline \& software,''
{\it IBM Research Report RC-19324,} 127 pp., December 1993.

\item
G. J. Chaitin,
``Randomness and complexity in pure mathematics,''
{\it International Journal of Bifurcation and Chaos\/}
{\bf 4} (1994), pp.\ 3--15.

\item
G. J. Chaitin,
``Responses to `Theoretical Mathematics\ldots',''
{\it Bulletin of the American Mathematical Society\/}
{\bf 30} (1994), pp.\ 181--182.

\item
G. J. Chaitin,
``The limits of mathematics (in C),''
{\it IBM Research Report RC-19553,} 68 pp., May 1994.

\end{itemize}
{\bf See Also:}
\begin{itemize}

\item
M. Davis,
``What is a computation?,'' in
L.A. Steen,
{\it Mathematics Today,}
Springer-Verlag, 1978.

\item
R. Rucker,
{\it Infinity and the Mind,}
Birkh\"auser, 1982.

\item
T. Tymoczko,
{\it New Directions in the Philosophy of Mathematics,}
Birkh\"auser, 1986.

\item
R. Rucker,
{\it Mind Tools,}
Houghton Mifflin, 1987.

\item
H.R. Pagels,
{\it The Dreams of Reason,}
Simon \& Schuster, 1988.

\item
D. Berlinski,
{\it Black Mischief,}
Harcourt Brace Jovanovich, 1988.

\item
R. Herken,
{\it The Universal Turing Machine,}
Oxford University Press, 1988.

\item
I. Stewart,
{\it Game, Set \& Math,}
Blackwell, 1989.

\item
G.S. Boolos and R.C. Jeffrey,
{\it Computability and Logic,}
third edition,
Cambridge University Press, 1989.

\item
J. Ford, ``What is chaos?,'' in
P. Davies,
{\it The New Physics,}
Cambridge University Press, 1989.

\item
J.L. Casti,
{\it Paradigms Lost,}
Morrow, 1989.

\item
G. Nicolis and I. Prigogine,
{\it Exploring Complexity,}
Freeman, 1989.

\item
J.L. Casti,
{\it Searching for Certainty,}
Morrow, 1990.

\item
B.-O. K\"uppers,
{\it Information and the Origin of Life,}
MIT Press, 1990.

\item
J.A. Paulos,
{\it Beyond Numeracy,}
Knopf, 1991.

\item
L. Brisson and
F.W. Meyerstein,
{\it Inventer L'Univers,}
Les Belles Lettres, 1991.
(English edition in press)

\item
J.D. Barrow,
{\it Theories of Everything,}
Oxford University Press, 1991.

\item
D. Ruelle,
{\it Chance and Chaos,}
Princeton University Press, 1991.

\item
T. N{\o}rretranders,
{\it M{\ae}rk Verden,}
Gyldendal, 1991.

\item
M. Gardner,
{\it Fractal Music, Hypercards and More,}
Freeman, 1992.

\item
P. Davies,
{\it The Mind of God,}
Simon \& Schuster, 1992.

\item
J.D. Barrow,
{\it Pi in the Sky,}
Oxford University Press, 1992.

\item
N. Hall,
{\it The New Scientist Guide to Chaos,}
Penguin, 1992.

\item
H.-C. Reichel and E. Prat del la Riba,
{\it Naturwissenschaft und Weltbild,}
H\"older-Pichler-Tempsky, 1992.

\item
I. Stewart,
{\it The Problems of Mathematics,}
Oxford University Press, 1992.

\item
A.K. Dewdney,
{\it The New Turing Omnibus,}
Freeman, 1993.

\item
A.B. \c{C}ambel,
{\it Applied Chaos Theory,}
Academic Press, 1993.

\item
K. Svozil,
{\it Randomness \& Undecidability in Physics,}
World Scientific, 1993.

\item
J.L. Casti,
{\it Complexification,}
HarperCollins, 1994.

\item
M. Gell-Mann,
{\it The Quark and the Jaguar,}
Freeman, 1994.

\item
T. N{\o}rretranders,
{\it Verden Vokser,}
Aschehdoug, 1994.

\item
S. Wolfram,
{\it Cellular Automata and Complexity,}
Addison-Wesley, 1994.

\item
C. Calude,
{\it Information and Randomness,}
Springer-Verlag, in press.

\end{itemize}

\end{document}